# A simulation study to check the accuracy of approximating averages of ratios using ratios of averages

J.M. van Zyl

ABSTRACT

For a number of researchers a number of publications for each author is simulated using the zeta distribution and then for each publication a number of citations per publication simulated. Bootstrap confidence intervals indicate that the difference between the average of ratios and the ratio of averages are not significant, and there are no significant differences in the distributions in realistic problems when using the two-sample Kolmogorov-Smirnov test to compare distributions. It was found that the log-logistic distribution which is a general form for the ratio of two correlated Pareto random variables, give a good fit to the estimated ratios.

## 1. Introduction

Summary citation data are often available but not summary results for individual researchers. An example is where Scopus provides a huge data base of research output of countries in terms of totals per subject field (SJR — SCImago Journal & Country Rank, 2007). This study is concerned with a common problem in bibliometrics, whether the ratio of averages (or totals) can be used as a proxy for the average of ratios of the individuals. Using resampling methods (i.e., bootstrap) in order to obtain confidence intervals it is shown that the difference between the average of ratios and the ratio of average is not statistically significant. The ratio of averages can be used as a proxy for the average. An application is where totals of publications and citations are available for each of a group of institutions, if the ratio of these totals can be used as a proxy for the average of individual ratios to do a ranking of the average number of citations per researcher of these institutions.

Egghe (2012) derived certain mathematical results concerning relationships between averages of ratios (AoR) and ratios of averages (RoA). He proved that the mean AoR and RoA are equal if the correlation between the ratios and the denominator used to calculate the ratios is zero. If the data of individuals and not totals are available, this result can be used, but in practice often only totals are available and a correlation cannot be calculated. In the simulation study the correlation between the ratio citations/publications and publications were found to be very close to zero and decreasing as the sample size increases.

Larivière and Gingras (2011) conducted a thorough study and tested equality of the medians of AoR and RoA using the Wilcoxon signed-rank test. For symmetric distributions medians and averages are equal, but not in general for skewed distributions. The distribution of differences of the AoR's and RoA,s and of the logs of AoR and RoA are not symmetric in the cases investigated in this study, which implies that the medians and means may differ. Conclusions based on tests for medians may thus not be valid for the means.

Bootstrap methods where no distributional assumptions are made were used in the simulation study to calculate confidence intervals for the mean difference between AoR and RoA. It was found that there are no significant differences between the means of AoR and RoA. Also for realistic parameters of the distribution of the number of publications, the distribution of AoR and RoA did not differ significantly, when using the Kolmogorov-Smirnov non-parametric two-sample test to test for equality of distributions.

In large samples, testing for equality of means or medians, will be significant even if the difference is negligible from a practical viewpoint. For example when looking at two means in the order of say size ten, a difference of say 0.05 in practice, is not unrealistic, but if the sample is large enough this difference will often be found to be significant, since the standard test is for exact equality.

In order to investigate results where citation statistics are involved, the simulation should be according to the distributional laws involved in citation data. There is still much research and development of models in this field, but the discrete

Pareto distribution or zeta distribution to model the number of papers generated by individual authors is often used. The zeta density is

$$p(k) = k^{-\gamma} / \zeta(\gamma), \quad k = 0,1,2,3..., \qquad (1)$$

where $\zeta(\gamma)$ is the Riemann zeta function which is finite for $\gamma > 1.0$ and is defined as $\zeta(\gamma) = \sum_{k=1}^{\infty} k^{-\gamma}$. Estimated values of $\gamma$ in the region of three are often found when using real data. Applications of this distribution can be found in the papers by Goldstein, Morris and Yen (2004), Redner (1998).

A typical set of citation data is that compiled by H. Small and D. Pendlebury of the data of the Institute for Scientific Information (ISI) which covers all publications from ISI catalogued journals that were published in 1981 and cited during the period January 1981 - June 1997. The frequency for each number of citations is also given, thus an empirical density estimate for the distribution of citations. This set of data is available on the website of Sidney Redner and also described in the papers of Redner (1998) and Peterson, Pressè, Dill (2010). It comprises the number of citations in that year of 344589 papers.

In section 2 a simulation study was conducted to investigate the differences between average AoR and RoA. For a specific number of researchers, a number of publications were simulated for each researcher using the zeta distribution and then according to the empirical density of citations a number of citations was simulated for each publication of an author. Thus using the probability calculated from the ISI data, a citation number was simulated for a specific publication.

The average over citations per publication per researcher was calculated and also the average of the ratio of total number of citations to total number of publications for this group of researchers. This was repeated a 1000 times and these values were used to construct confidence intervals for the differences between RoA and AoR.

The result shows that the RoA is a good approximation of AoR even in small samples. It should be kept in mind that both of these are random variables and even if one mean is less than the other theoretically, if the means are close the smaller mean could yield a larger sample mean for a specific sample. With this restriction kept in mind, it can be seen that the use of RoA is a very good approximation of AoR.

Within reasonable bound one can assume that similar patterns would be found in other sets of data, although the parameters of the distribution of citations and number of publications might be different.

## 2. Distribution of the ratios

A possible distribution would be the ratio of two correlated Pareto random variables. This was tested and found to give good results when fitted to observed ratio, RoA's and AoR's. There are more than one version of a bivariate Pareto distribution (Mardia, 1962), Arnold (1983). Let X and Y be two dependent Pareto distributed random variables and consider the bivariate distribution of the two variables

$$f(x,y) = [\alpha(1+\alpha)/k_1 k_2](1 + x/k_1 + y/k_2)^{-\alpha-2}, \ k_1, k_2, \alpha > 0.$$

It can be shown, see for example Markovich (2009), that the ratio $R=X/Y$ has the cdf

$F_R(x) = 1 - k_1/(xk_2 + 1)$, which leads to a density of the form

$$\begin{aligned}f_R(x) &= k_1^2/(xk_1 + k_2)^2 \\ &= (k_1^2/k_2^2)/(1 + x(k_1/k_2))^2.\end{aligned} \qquad (2)$$

It can be seen that this is a generalized Pareto distribution (GPD) with index equal to 1. This specific case of the GPD is a special case of the log-logistic density, which is more heavy-tailed that the log-normal. Various densities were tested on different sets of the simulated ratios and for example the log-normal yields good results in some cases, but the log-logistic density gave the most consistent best results when fitted to the ratios. There is a theoretical motivation as explained above, since it is a general form of the ratio of two correlated Pareto random variables. Theoretically the mean of the log-logistic is finite for $\alpha > 0$, and is equal to $E(x) = \pi\beta/\alpha \sin(\pi/\alpha)$.

Let $\beta = k_2/k_1$ and $\alpha > 0$, then the log-logistic density with scale parameter $\beta$ and shape parameter $\alpha > 0$ is

$$f(x) = (\alpha/\beta)(x/\beta)^{\alpha-1}/(1+(x/\beta)^\alpha)^2. \qquad (3)$$

The distribution function is

$$F(x) = (x/\beta)^\alpha/(1+(x/\beta)^\alpha), \ x \geq 0.$$

Some authors and software use a notation which is in terms of the corresponding logistic density, which is the distribution of *log(x)* and of the form:

$$f(x) = \alpha e^{\alpha(x-\log(\beta))}/(1+e^{\alpha(x-\log(\beta))})^{-2}, \ -\infty < x < \infty. \qquad (4)$$

and for example Matlab give estimates for $\log(\beta), 1/\alpha$. When tested on the ratios for example on simulated samples with $\gamma = 3.5$, estimates of $\alpha$ in the region of 4.5 were found. In figure 1 and 2 in section 3 examples are shown of fitted and observed observations.

## 3. Simulation study

For say *n* researchers, a number of publications are simulated for each researcher according to the probabilities of the zeta distribution (1). The distribution was truncated at 5000 publications. This was done because of the slow convergence of the probabilities to zero, and it was checked that for example the mean of the truncated distribution is almost exactly equal to the theoretical distribution without truncation. The normalizing constant is $C = 1/\sum_{j=1}^{5000}[k^{-\gamma}/\zeta(\gamma)]$ and the probability for k publications is

$$p(k) = Ck^{-\gamma}/\zeta(\gamma). \qquad (2)$$

It can be noted that the theoretical mean of the zeta distribution is $\zeta(\gamma-1)/\zeta(\gamma), \quad \gamma > 2$, and the mean using the truncated distribution with $\gamma = 3$ is 1.3683 compared to the theoretical mean of 1.3684, indicating a close approximation.

The number of citations per publications was calculated according to the observed values. For example there were 70836 of 783339 publications with one citation, and the probability for a publication to have one citation is estimated as 0.0904=70636 / 783336.

For a specific n, value of $\gamma$, m=1000 publications and citation results were generated for each of the n researchers, the average of ratios and the ratio of averages were calculated. A 95% bootstrap confidence interval (CI) for the difference between mean RoA and AoR was calculated using 500 simulated estimated ratios of averages and averages of ratios. The difference between the two averages was tested and not normally distributed. It can be mentioned that the difference of the logs give a more symmetric distribution, but also not normally distributed.

Using the data of 50 researchers with $\gamma = 3$ in the zeta density, the averages over 1000 simulations of the AoR and RoA both differ by less than 0.1 from 8.5733 which is the average number of citations per publication in the ISI data.

The results for $n = 25, 50, 250, 1000$ and $\gamma = 2.0, 3.0, 3.5$ is given in tables 1 to 3.

It should be noted that theoretically for $\gamma = 2$ the mean is not finite although finite for the truncated distribution. More realistic values of $\gamma$ is in the region of three and larger as was found for example by Redner (1998). The Kolmogorov-Smirnov test was performed each time and in the cases investigated with $\gamma = 3.0, 3.5$, the hypothesis of equal distribution could not be rejected.

| $\gamma = 2.0$ | Mean AoR | Mean RoA | Lower bound, 95% Bootstrap CI | Upper bound, 95% Bootstrap CI | K-S test p-value |
|---|---|---|---|---|---|
| n=25 | 8.5706 | 8.6511 | -0.3330 | 0.2198 | 0.0000 |
| n=50 | 8.5025 | 8.5418 | -0.2315 | 0.1596 | 0.0000 |
| n=250 | 8.6170 | 8.6363 | -0.1038 | 0.0692 | 0.0000 |
| n=1000 | 8.5882 | 8.5844 | -0.0394 | 0.0546 | 0.0000 |

Table 1. Summary statistics of m=1000 simulations for the AoR and RoA of citations per publication, using n=1000 simulated samples and parameters. Bootstrap CI for difference in means included. The hypothesis of equal distributions rejected if p < 0.05.

| $\gamma = 3.0$ | Mean AoR | Mean RoA | Lower bound, 95% Bootstrap CI | Upper bound, 95% Bootstrap CI | K-S test p-value |
|---|---|---|---|---|---|
| n=25 | 8.3142 | 8.3460 | -0.1400 | 0.0804 | 0.7530 |
| n=50 | 8.5059 | 8.5497 | -0.1572 | 0.0487 | 0.2816 |
| n=250 | 8.5140 | 8.5221 | -0.0568 | 0.0412 | 0.2575 |
| n=1000 | 8.5590 | 8.5502 | -0.0159 | 0.0335 | 0.3344 |

Table 2. Summary statistics of m=1000 simulations for the AoR and RoA of citations per publication, using n=1000 simulated samples and parameters. Bootstrap CI for difference in means included. The hypothesis of equal distributions rejected if p < 0.05.

| $\gamma = 3.5$ | Mean AoR | Mean RoA | Lower bound, 95% Bootstrap CI for difference in means | Upper bound, 95% Bootstrap CI for difference in means | K-S test p-value |
|---|---|---|---|---|---|
| n=25 | 8.5485 | 8.4946 | -0.0525 | 0.1543 | 0.7888 |
| n=50 | 8.4118 | 8.4837 | -0.1496 | 0.0003 | 0.9519 |
| n=250 | 8.5866 | 8.6160 | -0.1164 | 0.0104 | 0.7161 |
| n=1000 | 8.5525 | 8.5597 | -0.0335 | 0.0127 | 0.7530 |

Table 3. Summary statistics of m=1000 simulations for the AoR and RoA of citations per publication, using n=1000 simulated samples and parameters. Bootstrap CI for difference in means included. The hypothesis of equal distributions rejected if $p < 0.05$.

In the following table with $\gamma = 3.5$, the correlations were calculated to investigate the result of Egghe (2012) concerning the regression between the ratios and denominators. Thus for a specific $n$, 1000 correlations between citations/publications and publications were calculated using $n$ pairs each time. It was found that the correlation is on average very close to zero and decreasing as the sample becomes larger. The distribution of the observed correlations is skewed to the right in small samples. These results indicate that for large sample sizes AoR and RoA are approximately equal.

| $\gamma = 3.5$ | Mean correlation | Variance | Minimum | Maximum |
|---|---|---|---|---|
| n=25 | 0.0142 | 0.0263 | -0.2550 | 0.7945 |
| n=50 | 0.0095 | 0.0127 | -0.1825 | 0.7885 |
| n=250 | 0.0018 | 0.0015 | -0.0780 | 0.2228 |
| n=1000 | 0.0013 | 0.0003 | -0.0426 | 0.1096 |

Table 5. Summary results of m=1000 simulations for the correlation between citations/publications and publications.

A histogram of the 1000 observed AoR's is given in figure 1 and a histogram of the observed RoA's given in figure 2. It was with 50 researchers and $\gamma = 3.5$. It can be noted that in 569 of the 1000 samples, AoR was larger than RoA. This total fluctuates as one would expect from a random variable, but mostly in the region of half of the cases.

The observed mean of AoR for the 1000 simulated values was 8.5687 and the RoA mean was estimated as 8.6437. The 95% bootstrap confidence interval for the difference between these two means is (-0.1662, 0.0270) and the p-value for the two-sample Kolmogorov-Smirnov test is 0.8228.

A log-logistic distribution was fitted toe the RoA's and the estimated $\alpha = 4.7054$, $\beta = 7.8930$. In figure 1, a histogram with the fitted density is shown.

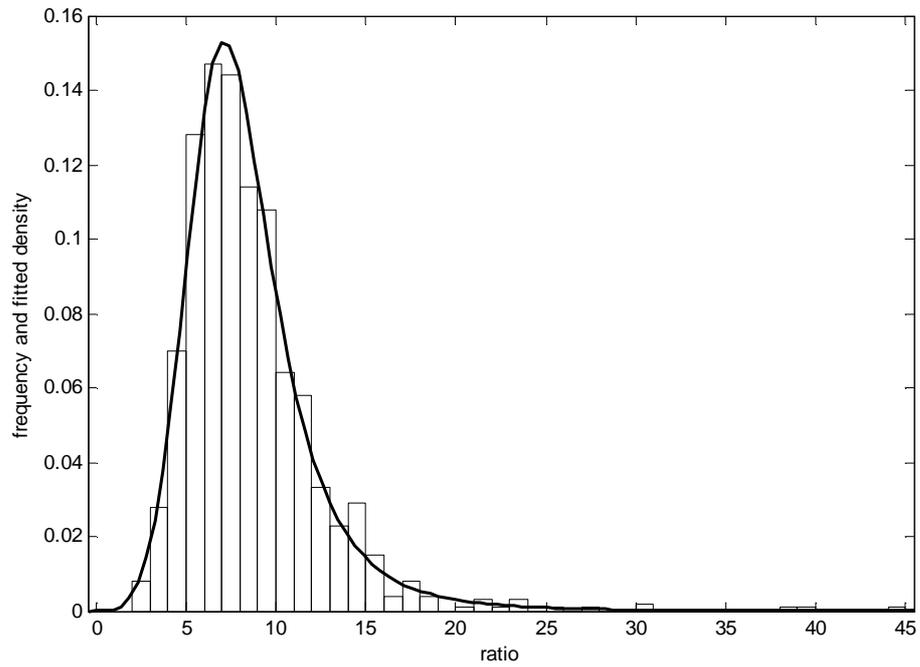

Figure 1. A histogram of 1000 AoR's and fitted log-logistic density calculated using 50 researchers and $\gamma = 3.5$.

The estimated parameters when the log-logistic was fitted to the 1000 observed RoA's are $\alpha = 4.7059$, $\beta = 7.8932$.

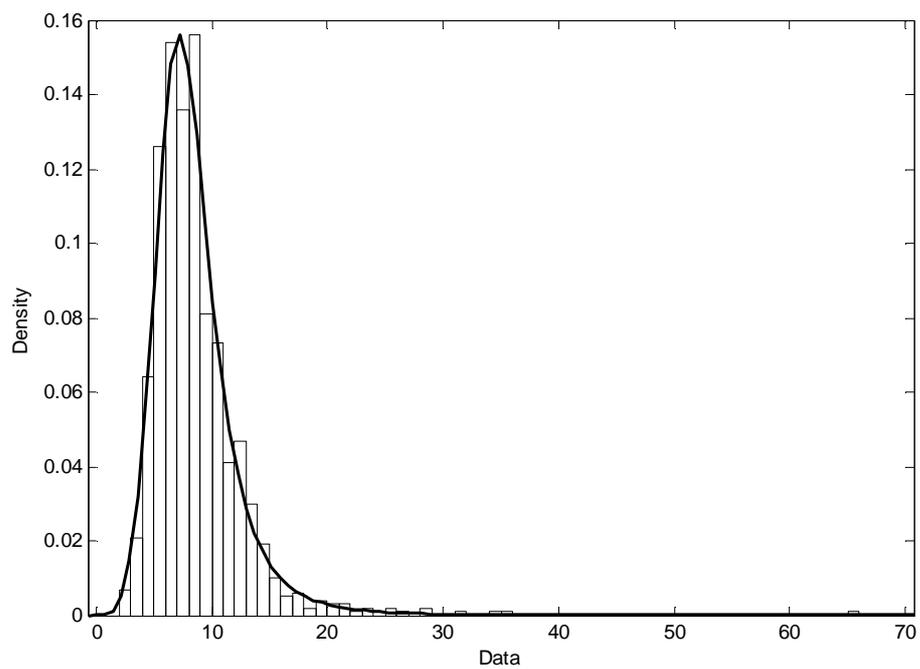

Figure 2. A histogram of 1000 RoA's and fitted log-logistic density calculated using 50 researchers and $\gamma = 3.5$.

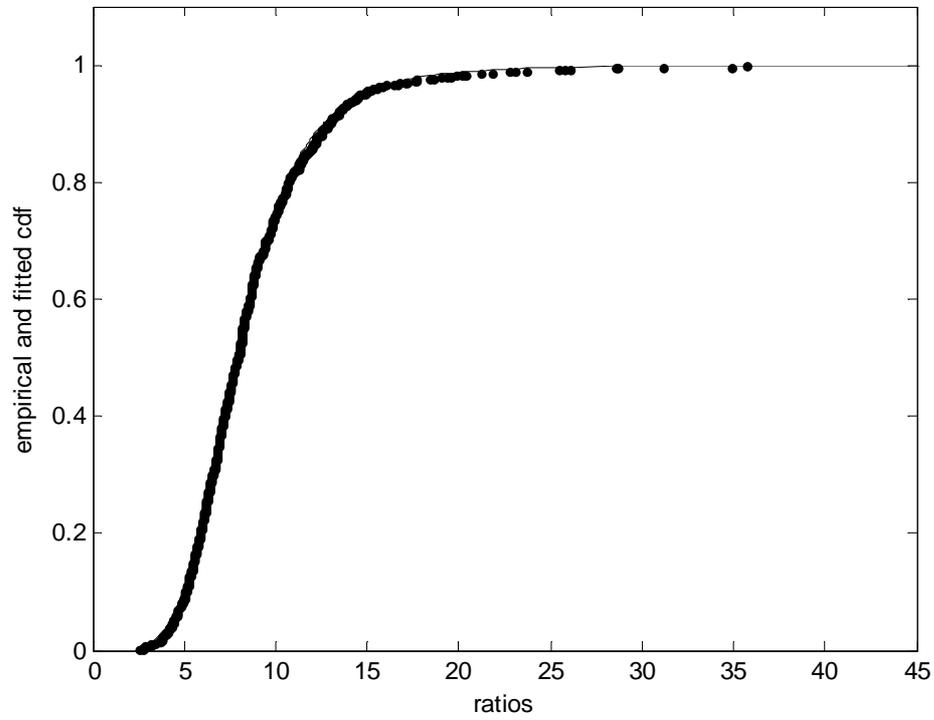

Figure3. Empirical and fitted log-logistic distribution function for 1000 RoA's using 50 researchers and $\gamma = 3.5$.

## 4. An application to country citation ranks

Scimago (2007) give research outputs of countries for the period 1996 – 2011. The results is given in terms of totals, thus total number of citable documents and total number of citations. If the totals are used to approximate the average number of citations per researcher, the result is given in table 4, showing that some of the smaller countries with respect to population produce high quality researcher if measured by citations. These averages are calculated over all subject fields.

| Rank with respect to research output | Country | Citations per document |
|---|---|---|
| 17 | Switzerland | 22.46 |
| 24 | Denmark | 21.17 |
| 14 | Netherlands | 20.82 |
| 1 | United States | 20.51 |
| 18 | Sweden | 19.78 |
| 25 | Finland | 18.28 |
| 7 | Canada | 18.19 |
| 3 | United Kingdom | 18.03 |
| 21 | Belgium | 17.81 |

Table4. Ranking of top 10 countries in terms of average citations per researcher together with ranking in terms of research output.

The sample sizes on which the results were calculated is very large and as can be seen, these ratios of averages give a reliable ranking, keeping in mind that these are random variables.

## 4. Conclusions

Based on the simulation study, it can be seen that using summary data and RoA's to approximate averages of individual ratios, is quite reliable even in small samples, for the number of publications and citations generated according to the laws governing the distribution of publications and citations.

This is a specific sample, but one can expect very similar results if publications and citations follow approximately the same type of distributions.